# Judging a socially assistive robot (SAR) by its cover; The effect of body structure, outline, and color on users' perception

Ela Liberman-Pincu*

Ben-Gurion University of the Negev, elapin@post.bgu.ac.il

Yisrael Parmet

Ben-Gurion University of the Negev, iparmet@bgu.ac.il

Tal Oron-Gilad

Ben-Gurion University of the Negev, orontal@bgu.ac.il

Socially assistive robots (SARs) aim to provide assistance through social interaction. Previous studies contributed to understanding users` perceptions and preferences regarding existing commercially available SARs. Yet, very few studies regarding SARs' appearance used designated SAR designs, and even fewer evaluated isolated visual qualities (VQ). In this work, we aim to assess the effect of isolated VQs systematically. To achieve this, we first conducted market research and deconstructed the VQs attributed to SARs. Then, a reconstruction of body structure, outline, and color scheme was done, resulting in the creation of 30 new SAR models that differ in their VQs, allowing us to isolate one character at a time. We used these new designs to evaluate users' preferences and perceptions in two empirical studies. Our empirical findings link VQs with perceptions of SAR characteristics. These can lead to forming guidelines for the industrial design processes of new SARs to match user expectations.

## 1 INTRODUCTION

The COVID19 crisis affected the lives of many. People chose or were directed to stay home during lockouts, except for essential activities. This new reality pointed toward a growing need for Socially Assistive Robots (SARs) for various contexts, functions, and users, in the para-medical field, for domestic uses, supporting elderly and disabled people, or helping in a variety of activities with children. SARs can be defined as the intersection of assistive robotics (AR) and socially interactive robotics (SIR) as they provide assistance through social interaction [1]. SARs may provide motivation, encourage training, and support rehabilitation to various users using nonphysical interaction [2]. Since there may be many functional differences among SAR types, it is expected that user experience will vary with the context of use, functionality, user characteristics, and environmental conditions. However, as we present in our market research, SAR manufacturers often designate the same embodiment for diverse contexts (i.e., one design fits all). Moreover. Our literature review revealed a gap between designers' and manufacturers' perceptions and users' perceptions in the question of what the user needs and wants [3].

During a new SAR's development process, attention centers primarily on specifications that guarantee proper functionality and compliance with safety regulations. Unfortunately, technical and functional specifications usually do not fully consider user needs nor guarantee SAR acceptance or users' attachment to the SAR. Hence, though researchers have acknowledged the difficulties of integrating user requirements into robotic designs for many years, we argue that now, with the market growth, it must be done (SAR presence grew by 44% in 2019 and continues to grow; IFR, 2020) [4]. In the para-medical field, the adoption of SARs can significantly reduce the risk of infectious disease transmission to medical professionals by allowing them remote diagnosis, monitoring, and treatment of patients [5]. In the domestic arena, SARs can support people's daily tasks within or outside their home, with chores, monitoring, providing companionship and supporting older adults, attending to the needs of the disabled, or helping with children. Since SARs can be classified in more than one way, based on the task and environment they perform in [6], their level of autonomy [1], or proactive and motion abilities[7], the industrial design of a SAR must account for these functional differences, also considering the users, the expected user experience, and the environment. The physical form of the SAR can help users understand its nature and capabilities [8]. Its appearances can differ by its human-likeness, structure, color, dimensions, etc. Matching a robots' appearance to its task/s can improve users' acceptance [9] is a long agreed-upon statement, but how this can be done is less studied.

Industrial design is an essential part of the process that leads to the development of new products, devices, and objects that have not existed before. Industrial designers focus on the visual appearance, usability, and manufacturability of a product, to achieve a high-value product and enhance the user experience of end-users. During the design process, the design team must make numerous decisions regarding the product's aesthetics and how it is going to be used by its users. Designers create a message encoded in a product by its geometry, dimensions, textures, and other VQs, leading the user to cognitive, affective, and behavioral responses [10]. Designing a SAR brings additional variables to the equation, first related to human-robot interaction. As opposed to static objects, or other kinds of robots, SARs must interact with users, and the quality of the interaction (QoI) impacts the user experience [11]. Second, SARs use motion as non-verbal communication means beyond visual appearance [12,13]. Third, SARs operate in dynamic changing circumstances, which may impact the dynamic relationships with them and how they are perceived by users, even within domains.

Within the world of robotics, and SARs particularly, the suitable and desirable user perception may vary according to the robot's context of use and the dynamics of the interaction; therefore, we previously proposed two models of relationships to lead to a better selection of VQs [14]. Onnasch & Roesler (2020) [15] had linked morphology (anthropomorphic, zoomorphic, technical) to robot tasks and human roles, suggesting different Human-robot interaction (HRI) scenarios deserve other embodiments. Other studies pointed to the need to adjust the robot's display screen to its task, understanding the importance of establishing different user perceptions for different domains. Broadbent et al. (2013) [16] suggested using a human-like face display only in cases where the robot should be perceived as sociable and amiable and avoid it in other cases. Kalegina et al., 2018, [17] explored the effect of 17 different faces on participants' perception of the robot's characteristics and role ascription.

Yet, our literature review revealed that very few studies dealt with isolated visual features using designated purposely designed SARs rather than existing off-the-shelf SARs. Most of the visual appearance studies follow the uncanny valley theory and focus on anthropomorphism, ranging from androids to cartoon-like robots [18-22]. Except for a few studies (e.g., [23,24]), detailed design factors have rarely been investigated. None of the studies addressed the effects of body structure and inner proportions (for example, screens or LEDs dimension and location).

Our current work aims to evaluate the effect of SARs' VQs on user perceptions of the SAR's characteristics. To do so, we first deconstruct the VQs attributed to SARs as detailed in section 3. This stage is then followed by a *reconstruction* process. The reconstruction design ended with the creation of 30 new SARs that differ from one another by one VQ or more. These SAR design models allow us to isolate and evaluate each VQ. We then use these SAR models to assess users' preferences and perceptions in two different studies, as detailed in the following sections. To narrow down the application scope and robot type, we chose to focus on one type of SARs, a machine-like personal assistant for home use.

## 2 RELATED WORK

### 2.1 Social Assistive Robots design

Design plays an important role in creating trust in a robotic system. The visual aesthetics of a product form the user's initial judgments (first impressions). Based on visual information, consistent first impressions can be formed within the first 39 milliseconds. In most cases, the first impression affects human behavior, attitude, and relationship with the product in the mid-and long term [25]. Studies have found that even the color alone has a psychological effect on the user by triggering the human arousal system and affecting product perception and trust in the product [26,27]. Designers use the word desirability to describe how attractive the visibility of a product is and whether the product is perceived as "worth having or seeking" and as being useful, advantageous, or pleasing [28].

SARs designs can be generally categorized into human-like and androids, animal-like, and machine-like [29-31]. Studies have found that robots with a more human-like design were perceived as more intelligent, yet participants tend to prefer working with a less human-like social robot [32]. Studies conducted among the elderly population have found that the elderly prefer having a robot that looks like a familiar object for their home setting [33,34]. Anthropomorphic robots were found to be less socially acceptable compared to machine-like robots [35,36]. The dimensions of the SAR affect the way it is perceived; smaller SARs were found to be more attractive to older adults [34].

### 2.2 Product personality and Visual Qualities (VQs)

People tend to attribute personality to different objects using human-like characteristics such as credible, friendly, professional, etc. The product's VQs determine the user's first impression and contribute to the user's emotions, associations, and perception of the product [37]. The VQs of a product may express its innovativeness or, on the other hand, a perception of a familiar old friend and have four semantic functions: describing its purpose and the way it should be used, expressing its values and qualities, signaling the user to specific reaction and identifying the product's nature, origin, product range, etc. [37]. The product's human-like characteristics may help the users anticipate the product's behavior and capabilities [38]. The product's structure, color, and materials are different tools for the designer to achieve the desired look & feel and user experience to communicate the intended product character and affect the user's interaction with the product [39,40]. Earlier studies have found that people prefer products with a similar personality to their own, as these products may help them confirm and express themselves [38]. Earlier studies linked VQs to users' perceptions of different products. Demirbilek and Sener (2003) [37] explored evoking emotions such as happiness and joy using the product's proportions and colors. Perez-mata et al. (2016) [41] explored the effect of vases' geometric features, such as curves, vertical-horizontal aspect ratio, color, etc., on the user's perception of the vases as being beautiful, aggressive, expensive, dynamic, etc.

In the field of HRI, a robot's perceived personality and even its perceived gender affect the user's expectations and help define the interaction. Studies have found that even subtle gender cues integrating into a robot's embodiment can significantly affect the HRI and the perception of the robot's identity, behavior, and suitable stereotypical tasks [42,43].

## 3 DECONSTRUCTION OF ROBOTS' VISUAL QUALITIES -MARKET RESEARCH

We used the google images search engine to collect images of commercial robots using the search words: socially assistive robot (the search was conducted in May 2020). After collecting various images and videos of SAR models, we deconstructed them into building blocks by visual components. The deconstruction phase is similar to describing the robot's appearance to someone who can't see it; first, we describe the robots' general appearance, its figure (i.e., Is it human-like? Pet-like? How realistic it is?), structure, colors, dimensions, etc. Then we look at smaller details and the use of the SARs' main components in the design (i.e., how the designer used the screen or the wheels in the design). The deconstruction process revealed a variety of different elements in the designs of existing SARs. Together with the literature review, we identified recurring and noticeable elements that are likely to affect users' perceptions and behaviors at two VQs levels: a) general appearance and b) main components use, as detailed below and summed in Figure 1. Naturally, some exceptional robots couldn't fit one or more classifications (e.g., unique color or shape). Nevertheless, the following sections represent the most common VQs found in our survey.

### 3.1 General appearance- the packaging

**Anthropomorphism**. SARs' visual appearance can be classified based on their human likeness. For this, we classified it into three groups as common in the HRI literature: Human-like, Pet-like, and Unrecognized (including machine-like). Human-like and pet-like features an inner classification from very realistic to more abstract models (e.g., A highly detailed figure of a dog or an abstract representation of a donkey).

**Shape and Structure.** The shape of the body can be classified by five models of structure (Rectangle, A shape, V shape, hourglass, and diamond) based on the classification of human body shapes [44], as well as by a division into one whole unit as opposed to two separate units (considering the wheels and screen). Shapes affect people's perceptions in many design and engineering disciplines [45]. Squares are considered to represent stability, trust, honesty, order, conformity, security, equality, masculinity, maturity, balance, and stubbornness, while rectangles are considered to represent action, aggression, energy, sneakiness, conflict, tension, masculinity, and force. Circles are thought to represent completeness, gracefulness, playfulness, comforting, unity, protection, childlike, innocence, youth, energy, and femininity [46]. The orientation of the shape also affects the perception; pointing down triangles (V shape) are perceived as more threatening than pointing up triangles (A shape) [47,48].

**Outline**. We had identified two groups: rounded-edged and chamfered-edged. Previous studies have found that people tend to prefer rounded edges objects compared to more angular in various contexts [49-51]

**Dimensions**. The dimensions of the SAR affect the way it is perceived [34]. In general, SARs' dimensions can be placed on a scale from pet size to human size.

**Material.** Materials can be described by the material itself (e.g., plastic, synthetic fur) or on a scale from hard to soft materials. A study conducted to evaluate the effect of material choice on user perception [24] is a good example of isolating one feature. This study found that wood and fur rated higher on the aspect of warmth than plastic did. Fur was perceived as more discomforting than wood.

**Color**. Commercial robots can be found in various color schemes; still, in our market research, three main color schemes emerged:1) Mainly white with a small area of light grey and an additional light color 2) White with different shades of blue 3) Dark colors like black/ dark grey/ dark blue. Color is the most noticeable design characteristic at first sight [52]. Color is well known as arouser of various psychological reactions and feelings; the effect of color has been investigated in many disciplines: psychology, architecture, website design, etc. Studies on user preferences for colors have found inconsistent results for several reasons; first, color must be associated with a particular object [53]; second, color preferences are not universal and vary across individuals and groups [54]. Figure 1 illustrates our findings and presents a preliminary classification of general appearance components.

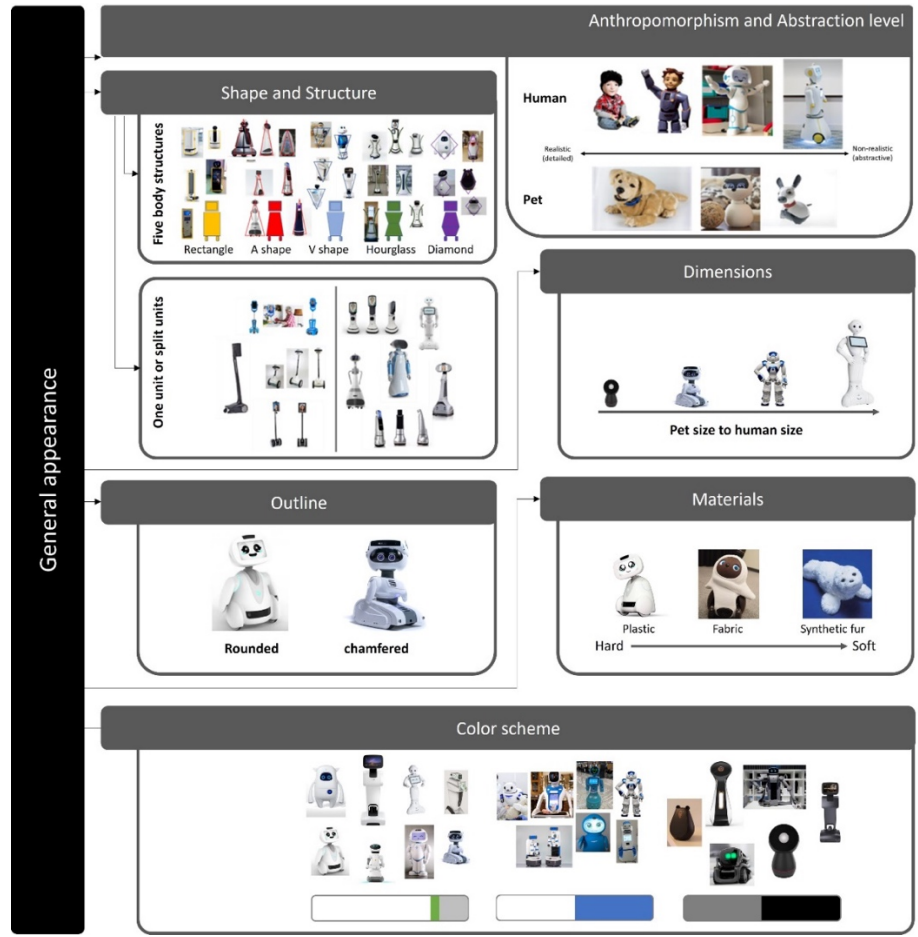

Figure 1: A classification of general appearance components of SARs.

### 3.2 The use of main components

Two recurring main components were found in most of the designs: a screen and wheels. The design team can use these components in different ways expressing varied characteristics. The **screen** is often used as the robot's head [55]; hence, it has the potential to be an essential tool to create the SAR's character; its orientation, location, and relative dimensions can affect users' perception. Previous studies investigated the effect of the human-like robot's head design and suggested that features such as the head shape, dimensions, head-to-body proportions, etc., heavily influence the users' perception of the robot's personality [23,56-58].

**Wheels** are an essential feature of mobile robots since they are easier to design and program than treads or legs, and in addition, in most cases, much cheaper [59]. The robots' wheels can be classified by their type (e.g., steering heel, Caster Wheel, etc.), the number of wheels, and the platform configuration [59, 60]. There is not much in the literature regarding wheels' design impact on users' perception, but it was found that the use of wheels rather than legs reduces human-likeness [61]. In our survey, we classified the use of the wheels in the design into three groups: hidden, shown, and emphasized, assuming these groups have the potential to affect the overall look of the SAR and the user's perception. Figure 2 illustrates our findings and presents a preliminary classification of the use of main components

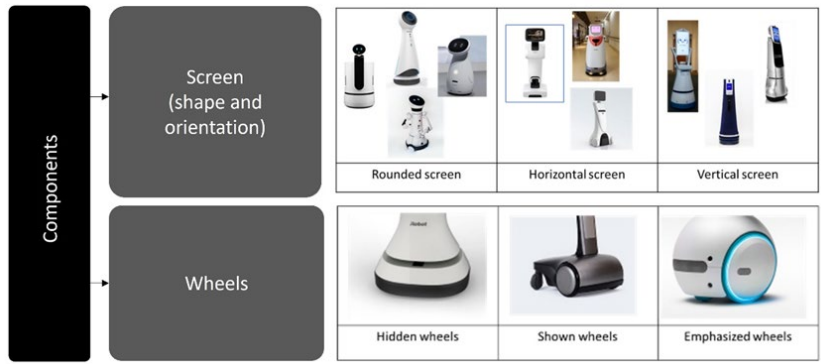

Figure 2: A classification of the use of main components of SARs, notably the screen and wheels.

### 3.3 Market research summary and conclusions

Our market research explored the design space of commercial SARs. Using a deconstruction process, we isolated common robots' VQs and created a preliminary taxonomy. Figure 3 summarizes the deconstruction phase and presents a preliminary taxonomy of VQs and the components to be later systematically evaluated. We then present two evaluation studies exploring three fundamental VQs: structure, outline, and color.

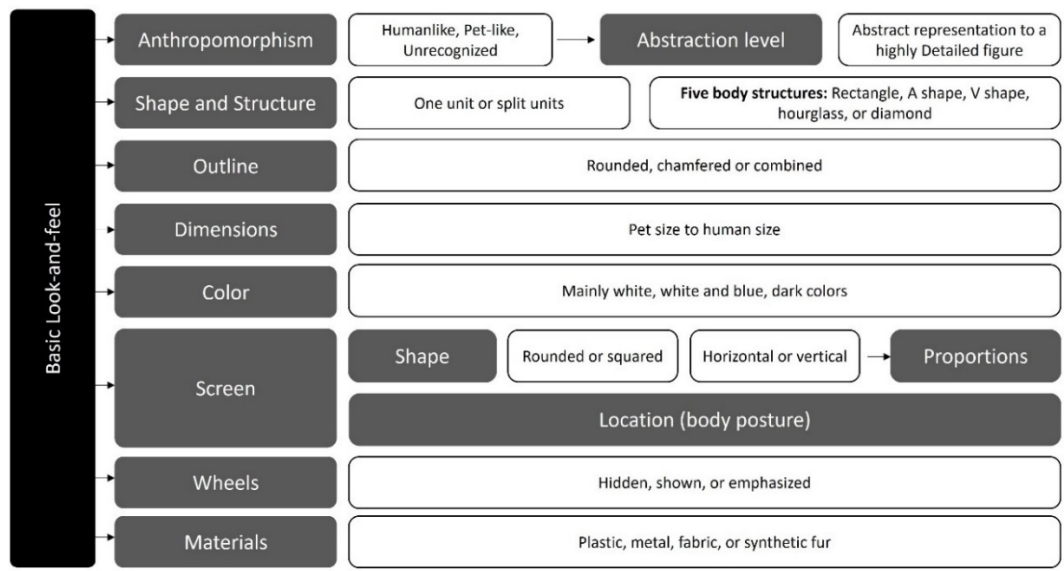

Figure 3: A preliminary taxonomy of Visual Qualities (VQs).

## 4 BUILDING BOXES FOR SARS, RECONSTRUCTION, AND EVALUATION STUDIES

Two studies, using online questionnaires, were run in parallel between August and October 2020, during the COVID-19 lockdown. To help participants relate to the SARs, we defined a context of a personal assistant to the elderly population in times of social isolation (during the COVID-19 outbreak) that can offer support with the user's daily tasks, communicating with family and friends, tracking vitals, and more. First, we describe the building boxes for the SAR designs and models, which are common to both studies. We then describe the measurement and assessment tools. Study 1 aimed to evaluate the effect of VQs on users' *perception* of the SAR, while Study 2 explored users' *preferences* when choosing the building boxes and assembling their own SAR. Three basic VQs were selected for the investigation: body shape, outline, and color (as detailed in the following section).

### 4.1 Study Design

#### 4.1.1 Creation of building boxes for new SARs models

We created our own 30 (5X2X3) abstractive SAR designs using the deconstructed VQs, allowing us to isolate one character at a time (shape, color, outline). A 3D detailed design using a CAD software model was created for each abstractive design. All models are the same height, having four small wheels that are used to support their movement around the house, and all contain a large horizontal screen allowing watching videos, conducting video chats, playing, etc. Table 1 details the selected VQs' levels; Figure 4 shows the building boxes and three examples of the self-constructed 3D robotic figures. These designs were used as the building boxes for our two studies to evaluate the effect of each specific factor on users' perceptions and preferences.

Table 1: The deconstructed VQs selected for the evaluation studies.

| Visual quality | Levels |
| --- | --- |
| Body structure | Rectangle, A-shape, V-shape, Hourglass, Diamond |
| Outline | Rounded, Chamfered |
| Color scheme | White and blue, Mainly white, Dark |

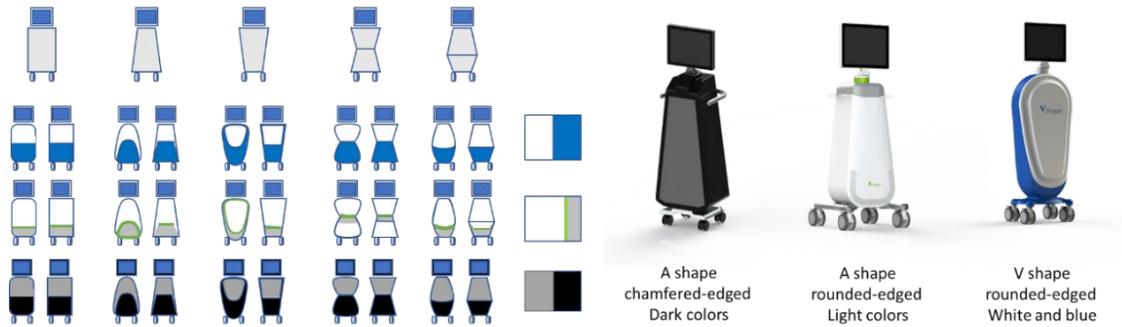

Figure 4: Left- Abstractive building boxes for SAR Visual Qualities based on our initial findings; Right- Three examples of 3D SAR figures.

### 4.1.2 Tools and materials

**Measuring attitude toward robots and technology**

Two standard questionnaires were used for measuring attitudes toward robots and technology: the Negative Attitudes towards Robots Scale (NARS) [62]; and the Technology Adoption Propensity (TAP) [63]. NARS consists of 14 statements classified into three subordinate scales, S1: "Negative Attitude toward Situations of Interaction with Robots", S2: "Negative Attitude toward Social Influence of Robots", and S3: "Negative Attitude toward Emotions in Interaction with Robots". TAP is a multiple-item scale developed to measure consumers' propensities to adopt new technologies. It uses a 14-item index that combines assessments of consumers' positive and negative attitudes towards technology.

**Measuring design perception**

Perception of visual appearance can be measured in different ways; ascribing product reaction words to different designs [28], using Semantic Differential scales (SD scales) where respondents are asked to indicate their position on a scale between two bipolar words or using Likert scales where subjects are asked to state their agreement with different statements. In the world of robotics, two common questionnaires are used for the evaluation of users' perceptions: Godspeed [64] and Robotic Social Attributes Scale (RoSAS) [65]. Godspeed measures five distinct dimensions: anthropomorphism, animacy, likeability, perceived intelligence, and perceived safety, using an SD scale. RoSAS defines three central evaluation factors: warmth, competence, and discomfort using a Likert scale. Based on these standard questionnaires and our preliminary studies, we create an alternative set of describing words that is more suitable for non-human-like robots to be used in our following studies: friendly, childish, innovative, threatening, intelligent, reliable, professional, massive, elegant, medical, and old fashioned. In our preliminary studies, we tried two different measurement approaches: (1) selecting suitable words out of a word bank; participants only chose the words they found to match their perception of the design- either yes or no, with no levels of agreement [66,67]. (2) Using a five-point Likert scale, participants were asked to react and provide their level of agreement for each describing word (e.g., how well does the adjective friendly describe the above robot? Very little, little, medium, high, or very high) [14]. We concluded that a Likert scale would be the most appropriate for our studies.

### 4.2 Research Questions

We aim to investigate the effect of VQs on users' perceptions and preferences of a SAR. The following questions were addressed: (1) What impact do the specific VQs, body structure, outline, and color scheme have on user perception? And (2) Are there shared preferences regarding the VQs of personal assistant robots for home use? Three research hypotheses were formulated and are detailed below.

Hypothesis 1 [H1] SARs' VQs affect user perception. Since there is not enough literature regarding the effect of isolated VQs in the field of robotic design, we based our hypothesis on previous studies in other design fields. Hsiao & Chen (2006) [68] found relationships between the object's shape features and user responses in three product categories: automobile, sofa, and kettle, representing large, medium, and small products.

[H1a] The body shape of the SAR affects the user's perception. That people respond to shapes is known in many design and engineering disciplines [45]. Squares are considered to represent stability, trust, honesty, order, conformity, security, equality, masculinity, maturity, balance, and stubbornness, while rectangles are considered to represent action, aggression, energy, sneakiness, conflict, tension, masculinity, and force. Circles are thought to represent completeness, gracefulness, playfulness, comforting, unity, protection, childlike, innocence, youth, energy, and femininity [46]. The orientation of the shape also affects the perception; pointing down triangles (V shape) are perceived as more threatening than pointing up triangles (A shape) [47,48]. Therefore, we assume that links between the robot's body structure and users' perception of its characterization and gender ascription will be found.

[H1b] The SAR's outline affects the user's perception. Previous studies have found that people prefer rounded edges to more angular objects in various contexts [49-51]. We assume that respondents will prefer rounded-edge robots and that a definite positive link will be found between rounded edges SARs and positive perception.

[H1c] The SAR color scheme affects user perception. Color is the most noticeable design characteristic at first sight [52]. Color is a well-known arouser of various psychological reactions and feelings; the effect of color has been investigated in many disciplines: psychology, architecture, website design, etc. Studies on user preferences for colors have found inconsistent results for several reasons; first, color must be associated with a certain object [53]; second, color preferences are not universal and vary across individuals and groups [54]. We expect to find different attributes to each color scheme and identify differences in the preferences according to the personal characteristics of the respondents.

[H1d] VQs that share similar perceptions will be selected together. The VQs of an object create its design language; we hypothesize that participants will choose VQs that share similar characteristics. This assumption is based on the author's previous findings in the medical design field, where a research market revealed three common design languages composed of different color schemes and outlines [69].

Hypothesis 2 [H2] personal characteristics affect design perception and preferences of SARs. Many fields of design use personas as part of the development process. Personas refer to fictional persons created based on a survey of users and represent a group of users with common characteristics, attitudes, and behaviors regarding their interaction with a particular product or service [70,71]. We hypothesize that a link between personas and design preferences will be found. Since, in the SAR field, most studies related to design preferences were using human-like and machine-like off-the-shelf robots, there is no clear way to evaluate VQs preferences but as a preferred choice out of several possible designs.

Hypothesis 3 [H3] positive perceptions will correlate with design preferences. Participants will tend to choose VQs related to positive perceptions (e.g., favor friendly over threatening).

### 4.3 Study 1: Evaluating the effect of visual qualities on user perception

#### 4.3.1 Aim and scope

In study 1, we used our self-designed SARs to evaluate adults' perception of three VQs in the context of a personal assistant to the elderly population using an online questionnaire.

#### 4.3.2 Method and online questionnaire design

Thirteen robots out of the 3twe designed robots were chosen. The selection was made so that comparisons could be analyzed for each VQ. All building boxes were maintained, but the number of alternatives was reduced to avoid lengthy questionnaires and the abandonment of participants. Figure 5 shows the 13 selected designs and illustrates the possible comparisons. For instance, the White and Blue color scheme can be compared across four body structures and so forth.

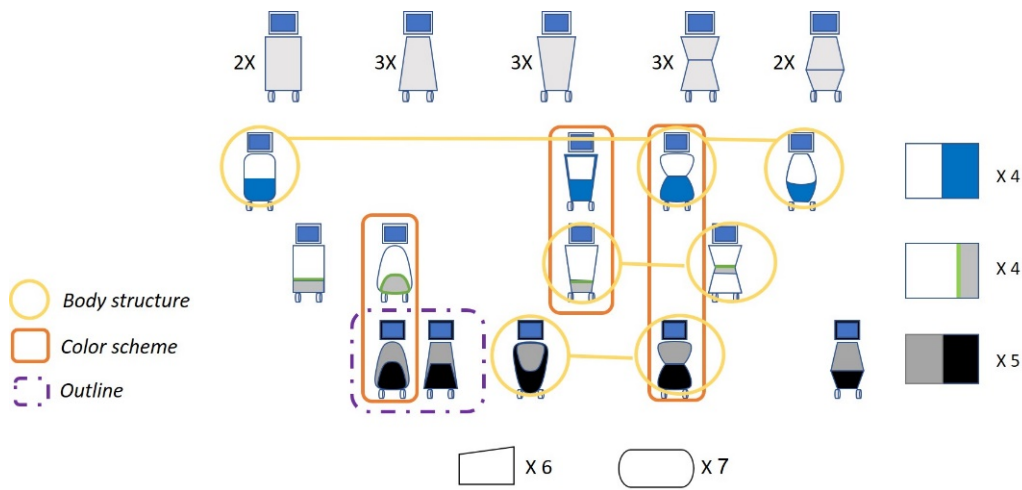

Figure 5: Selected designs for user perception questionnaires. 13 robots were rendered, and the lines illustrate how their distribution covered the possible combinations of body shape (Rectangle, A shape, V shape, Hourglass, and Diamond), color scheme (White and blue, Mainly white, and Dark), and outline (Chamfered or Rounded).

Using Qualtrics, we designed an online questionnaire constructed of two parts: (1) Personal characteristics – Participants were asked to fill in demographic information, TAP, and NARS questionnaires. (2) User perception questionnaire contained randomly selected seven out of the 13 robots. Showing one image at a time, robots were presented from the same angle and dimensions using similar backgrounds. Following each image, respondents were asked to describe their reactions towards the robot's design based on their first impression using a list of eleven describing words (friendly, childish, innovative, threatening, intelligent, reliable, professional, massive, elegant, medical, and old fashioned) and a five-point Likert scale. Additionally, they were asked if they could ascribe gender to the robots. Note that in Hebrew, the word robot is assigned to the male gender; hence there is an inherent bias toward ascribing the male gender to a robot. Images and describing words were shuffled within the questionnaire. Figure 6 shows the questionnaire structure and one example of a robot's image.

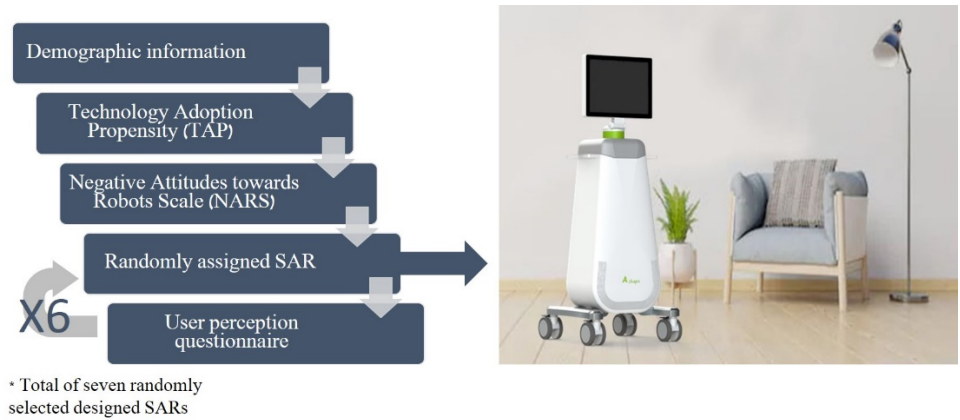

Figure 6: Right: workflow of the Perception questionnaire. Left: an example of a robotic image, as seen in the questionnaire (in this case, a mainly white, A-shaped rounded edges model).

### 4.3.3  *Participants*

The online questionnaire was distributed between August to October 2020 (in the midst of the COVID-19 restrictions), using social media (Facebook and WhatsApp) as a rolling snowball where participants were asked to further refer and post. In total, data from 160 adult respondents were collected (54% females and 46% males), 18 years old and older (M= 33.6 years, SD = 13.8).

### 4.3.4  *Analysis and Design.*

The original Likert scale in the questionnaire consisted of 5 ticks (levels of response). For some of the characteristics, there was less than 4 percent of the responses in the extreme cases (e.g., old fashioned, childish, threatening – tick 5) and equal to or less than 8 percent for the not at all (e.g., innovative, reliable, professional, wise – tick 1). We, therefore, decided to adopt a three tick Likert scale, negative, neutral, and positive, where ticks 4 and 5 and ticks 1 and 2 were merged. We are aware of arguments that this type of merging should not be done [72] but we argue that in the case of design, we need to take a more lenient approach, i.e., even if the effect is marginal or less dominant (i.e., middle ticks), it is better to include and consider it in the design than ignore.

A mixed within- between- design where each characteristic was defined as a dependent variable. A general linear mixed model (GLMM) analysis was conducted to determine which VQs and user characteristics significantly influence each perceived characteristic. Robot-id and participant-id were included in the model as random effects to account for individual differences among respondents and design differences among robots. Fixed effects were the participants' age and gender and the robots' VQ; structure, color, and outline.

### 4.3.5 Results

All three VQs were found to affect user perception of the SAR. Each VQ affected different characteristics perceptions; adjusting these VQs may help the designer present the SAR as being massive, friendly, threatening, elegant, childish, old-fashioned, medical, and innovative. Furthermore, we have found a significant effect on the perception of robot gender. Table 2 summarizes the results. Details are given in the following paragraphs. Estimated means are given in Tables 3-5, respectively, for structure, color, and outline. Ranges are from 1-3, with values closer to 3 indicating a higher probability that this robot would be assigned this characteristic.

Table 2: VQs' effect on self-designed SAR characteristics. Dark boxes represent significant effects.

|  | *Friendly* | *Childish* | *Innovative* | *Threatening* | *Old-fashioned* | *Massive* | *Elegant* | *Medical* | *Robot gender* |
|---|---|---|---|---|---|---|---|---|---|
| **Structure** | ■ |  | ■ | ■ |  | ■ | ■ |  | ■ |
| **Color** | ■ | ■ |  | ■ |  | ■ | ■ | ■ | ■ |
| **Outline** |  |  | ■ |  | ■ |  | ■ |  |  |

Significance level p<.05

**Structure.** The structure significantly affected user perception of five characteristics: friendly ($F(4,1054)=2.97$, $p=.019$), innovative ($F(4,1054)=4.4$, $p=.002$), threatening ($F(4,1054)=2.375$, $p=.05$), massive ($F(4,1054)=3.23$, $p=.012$), and elegant ($F(4,1054)=4.34$, $p=.002$). V-shape robots were perceived as more massive and threatening than all others, along with being the least friendly. But they were also perceived as innovative. Rectangle robots were perceived as the least innovative and elegant but also as not threatening. A-shape and hourglass robots were perceived as the friendliest. Hourglass robots were also perceived as the most elegant and innovative and the least massive. Figure 7 shows three robots (#3, #5, #9) sharing the same color and outline. Designs #3 and #5 were both perceived as elegant and innovative, but #3 was perceived as more massive.

Table 3: Predicted impact of the structure- Estimated means

|  | Rectangle (R) | | A Shape (AS) | | V Shape (VS) | | Hourglass (H) | | Diamond (D) | | Post-hoc comparisons with Bonferroni correction |
|---|---|---|---|---|---|---|---|---|---|---|---|
|  | Mean | SE | Mean | SE | Mean | SE | Mean | SE | Mean | SE |  |
| **Massive** | 2.1 | 0.065 | 2.2 | 0.055 | 2.2 | 0.051 | 2.0 | 0.052 | 2.1 | 0.066 | VS,AS>>H |
| **Friendly** | 1.8 | 0.066 | 2.0 | 0.056 | 1.7 | 0.052 | 2.0 | 0.053 | 1.9 | 0.068 | AS,H>>VS |
| **Threatening** | 1.3 | 0.054 | 1.4 | 0.046 | 1.5 | 0.043 | 1.3 | 0.043 | 1.3 | 0.055 | VS>>R,H,D |
| **Elegant** | 2.0 | 0.067 | 2.1 | 0.058 | 2.2 | 0.053 | 2.3 | 0.054 | 2.2 | 0.069 | H>>R |
| **Innovative** | 2.0 | 0.063 | 2.3 | 0.051 | 2.3 | 0.051 | 2.4 | 0.053 | 2.2 | 0.064 | H,VS,AS>>R |

(>) signifies p<.05,  (>>) signifies p<.01
*Only significant comparisons are presented in the right column

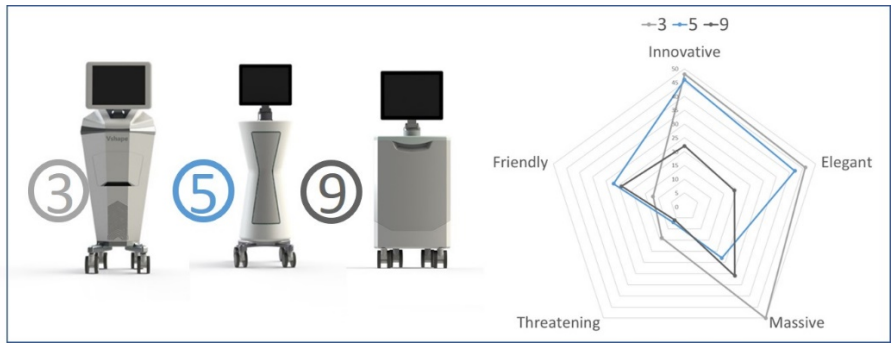

Figure 7: Structure's effect- a comparison of three robots sharing the same color and outline. Left – robot designs #3, #5, and #9. Right – a radar chart of participants' agreement to the describing words. The scale indicates the number of participants (out of 80) that indicated that the robot has a specific characteristic (i.e., rated it 4 or 5 on the Likert scale).

The body structure was found to affect the perception of gender; hourglass-shaped robots were perceived as more feminine than others (54% of the respondents addressed them as female), while V-shaped robots were perceived as most masculine (57% of the respondents addressed them as males) followed by the diamond shape (51%). Figure 8 shows the perception of gender ascribed by the participants.

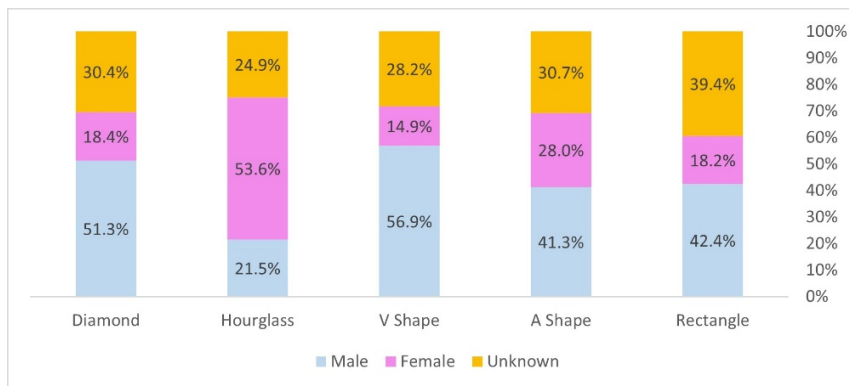

Figure 8: the perception of gender ascribed by the participants to the robot

**Color.** Color significantly affected user perception of six characteristics: massive ($F(2,1054)=4.89$, $p=.008$), friendly ($F(2,1054)=7.95$, $p<.001$), threatening ($F(2,1054)=13.37$, $p<.001$), elegant ($F(2,1054)=3.62$, $p=.027$), childish ($F(2,1054)=13.92$, $p<.001$), and medical ($F(2,1054)=29.49$, $p<.001$). Dark SARs were perceived as more massive and threatening than white and blue or mainly-white ones, and less friendly, but also elegant and less childish. The combination of white and blue SARs was found to be perceived as the friendliest color scheme. Mainly-white SARs were perceived as elegant and medical. Figure 9 demonstrates three sets of robot pairs sharing the same structure and outline.

Table 4: Predicted impact of color scheme - Estimated means

|  | White and Blue (WB) | | Mainly white (MW) | | Dark (D) | | Post-hoc comparisons with Bonferroni correction |
|---|---|---|---|---|---|---|---|
|  | Mean | SE | Mean | SE | Mean | SE |  |
| **Massive** | 2.0 | 0.046 | 2.1 | 0.048 | 2.2 | 0.043 | D>MW>>WB |
| **Friendly** | 2.0 | 0.047 | 1.9 | 0.048 | 1.7 | 0.044 | WB>MW>>D |
| **Threatening** | 1.2 | 0.039 | 1.3 | 0.04 | 1.5 | 0.036 | D>>MW>>WB |
| **Elegant** | 2.0 | 0.051 | 2.2 | 0.052 | 2.2 | 0.044 | MW,D>>WB |
| **Childish** | 1.4 | 0.035 | 1.3 | 0.035 | 1.2 | 0.03 | WB>MW>D |
| **Medical** | 2.2 | 0.046 | 2.3 | 0.046 | 1.9 | 0.04 | MW>WB>>D |

(>) signifies p<.05,  (>>) signifies p<.01
*Only significant comparisons are presented in the right column

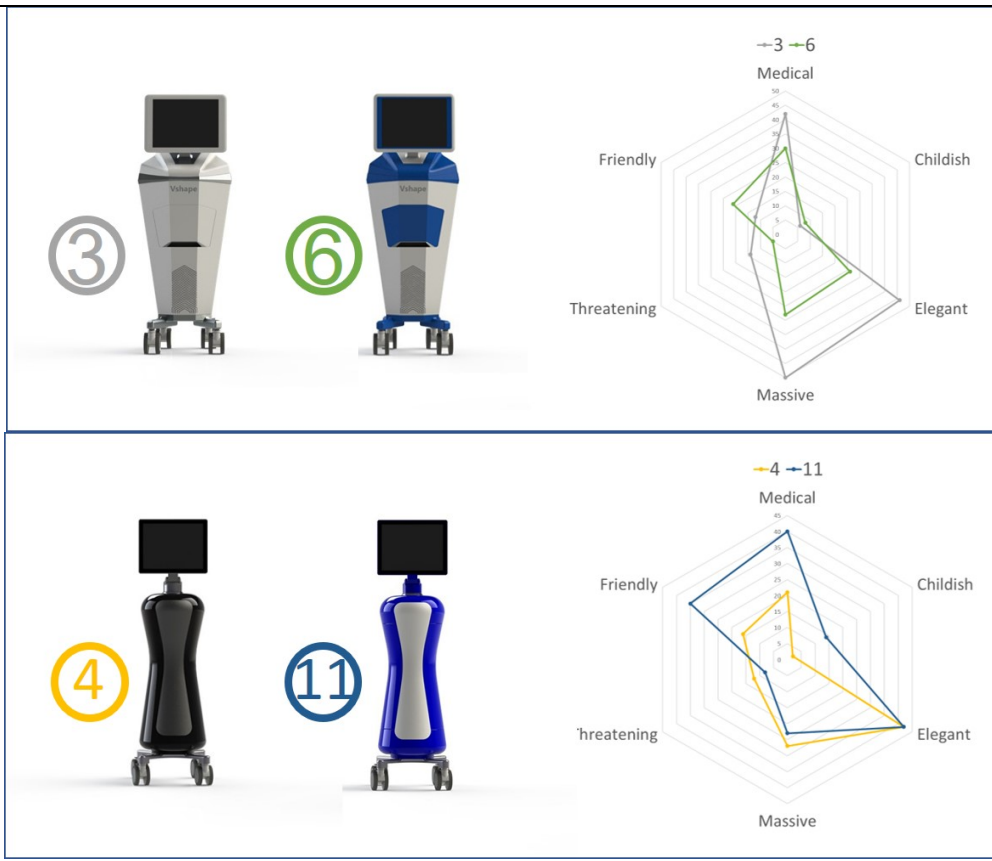

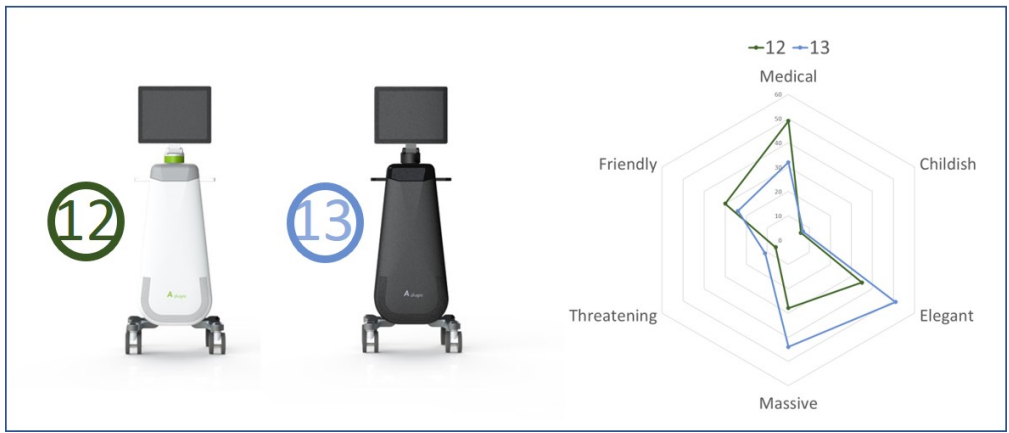

Figure 9: Color's effect- three sets *of* robot pairs sharing the same structure and outline. The radar charts present participants' agreement to the describing words. The scale indicates the number of participants (out of 80) that indicated that the robot has a specific characteristic (i.e., rated it 4 or 5 on the Likert scale).

.

In addition, color affects the perception of robots' gender. Participants tend to ascribe the male gender to robots in general, but the white color scheme may affect their tendency toward the female gender, as shown in Figure 10.

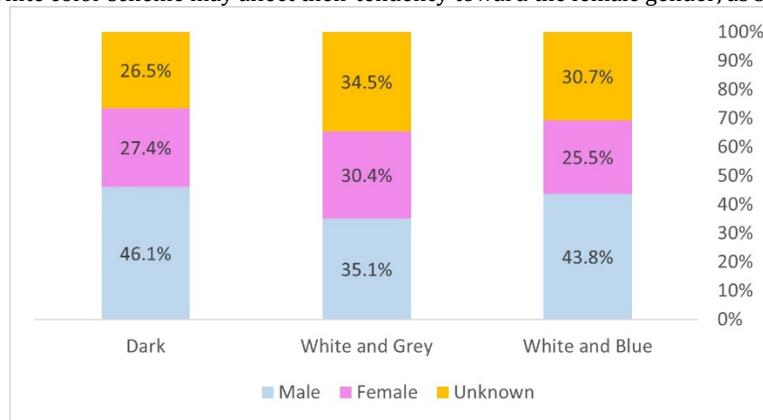

Figure 10: gender perception by color

**Outline.** Chamfered or rounded outlines affected the user's perception of three characteristics: elegant, innovative, and old-fashioned. Chamfered edges made the robot be perceived as old-fashioned (F(1,1054)=4.83, p=.028), less innovative (F(1,1054)=5.18, p=.023) and less elegant (F(1,1054)=8.21, p=.004). Figure 11 demonstrates the outline's effect by comparing two robots sharing the same structure and color. The chart presents differences in the perception of two more characteristics in addition to *Elegant, innovative,* and *Old-fashioned* that were found significant in the GLMM model: *Medical* and *Friendly*. It implies that a rounded outline is perceived as friendlier and more appropriate for the medical domain. Possibly, the outline is more effective than found in the model, and this was not reflected in the results due to few comparisons among the 13 chosen robotic designs.

Table 5: Predicted impact of the line- Estimated means

|  | Sharp (S) | | Rounded (R) | | Main effect |
| --- | --- | --- | --- | --- | --- |
|  | Mean | SE | Mean | SE |  |
| **Elegant** | 2.1 | 0.04 | 2.2 | 0.038 | R>>S |
| **Old Fashioned** | 1.6 | 0.035 | 1.5 | 0.032 | S>>R |
| **Innovative** | 2.2 | 0.039 | 2.3 | 0.038 | R>>S |

(>) signifies p<.05, (>>) signifies p<.01
*Only significant comparisons are presented in the right column

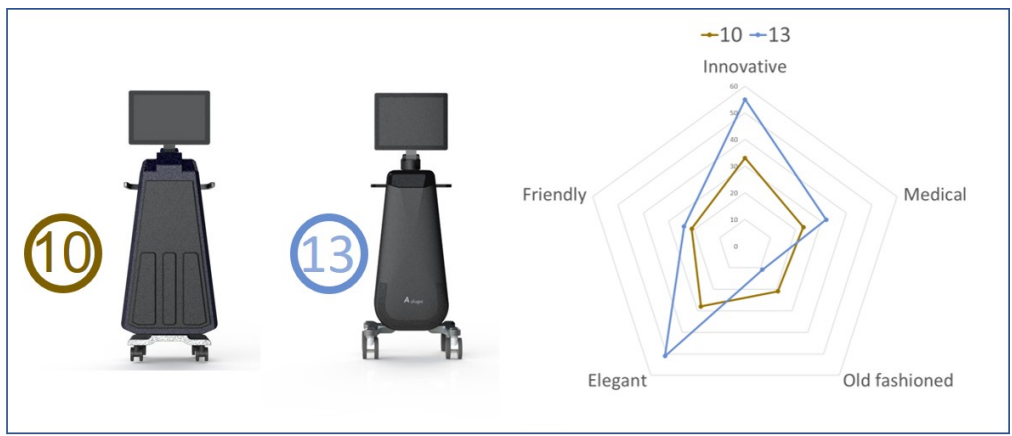

Figure 11: Outline's *effect*- a comparison of two robots sharing the same color and body struct. Left – robot designs #10, and #13. Right – a radar chart of participants' agreement to the describing words. The scale indicates the number of participants (out of 80) that indicated that the robot has a specific characteristic (i.e., rated it 4 or 5 on the Likert scale).

#### 4.3.6 *Summary*

We found that a designer can lead users to a perception of different robots' characteristics. These findings support new SAR design processes by providing insights regarding the effect of structure, color, and outline. For example, to achieve the perception of a friendly SAR, a designer should consider using A-shape or hourglass structures and avoid V-shape, choose light colors (a combination of white and blue is preferred), and avoid dark colors. Using rounded edges would make the SAR look more elegant and innovative, while chamfered edges would contribute to an old-fashioned device's perception. On the other hand, a V-shape structure and dark colors may be suitable for a more authoritative SAR; this appearance may be beneficial for tutoring robots and in cases where discipline is needed- for example- supervisor-robot, Inspection – porter robot, or a gatekeeper robot.

To confirm these findings, we explored the two specific cases presented above to verify that these combinations are indeed perceived as expected. We searched for the robots perceived as the friendliest and robots perceived as the most threatening; we expected to find that combining VQs that stimulate the same perception would increase the perception of the specific characteristic. Our study's average friendliness ranking was 2.7 out of 5: Two of the

13 robots ranked highest as friendly; robot #11, and robot #12, both with an average ranking of 3.1 out of 5. Their appearance matched our expectations; robot #11 is a white and blue, rounded, hourglass-shaped robot, and robot #12 is a white, rounded A-shaped robot. The average ranking of threatening was 2 out of 5: one of the 13 robots ranked highest as threatening; robot #7 had an average of 2.4 out of 5. Its appearance matched our expectations; it is a dark, rounded V-shaped robot. Note that not all 30 combinations were presented and evaluated (see figure 5), this may be the cause for resulting in three rounded robots for the two categories. For example, there was no option for a dark, chamfered V-shaped robot. Table 6 presents the two friendliest models and the most threatening one.

Table 6: Two friendliest models and the most threatening one.

| Perceived characteristic | Robot | Visual qualities | | |
|---|---|---|---|---|
| | | Structure | Outline | Color |
| **Friendliest robots** | Robot #11 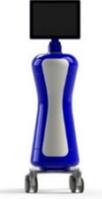 | Hourglass | Rounded | White and blue |
| | Robot #12 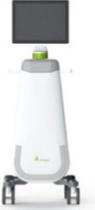 | A shape | Rounded | Mainly white |
| **Most threatening robot** | Robot #7 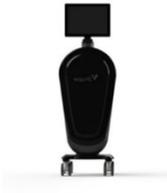 | V shape | Rounded | Dark |

### 4.4 Study 2: Exploring users' preferences

#### 4.4.1 Aim and scope

In study 2, we sought to explore users' preferences regarding body structure, outline (chamfered or rounded), color scheme, and the display screen - graphic user interface (GUI) of personal assistant robots for home use. We used our building boxes to let participants assemble their own personal SAR using an online questionnaire.

#### 4.4.2 Method and online questionnaire design

Using Qualtrics, we designed an online questionnaire constructed of three parts: (1) Personal characteristics – Participants were asked to fill in demographic information, TAP, and NARS questionnaires. (2) participants were asked to design their own personal assistant robot by choosing four elements that compose the robot's appearance: robot's body structure (V-shape, A-shape, diamond shape, hourglass, or rectangle), outline (rounded or sharp), color scheme (dark, mainly white, and blue and white), and finally the GUI utility icons or face display. (3) Participants watched a short animation presenting their final design. Figure 12 illustrates the preferences questionnaire structure.

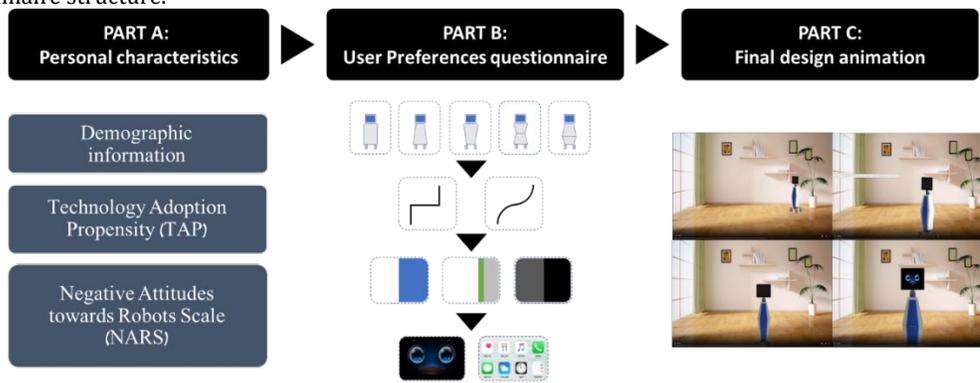

Figure 3: Preferences *questionnaire* structure. Participants assembled their own preferred structure for the SAR by selecting shape, outline, color scheme, and GUI.

#### 4.4.3 Participants

The online questionnaire was distributed between August to October 2020, using social media (Facebook and WhatsApp) as a rolling snowball where participants were asked to further refer and post. In total, data from 110 adult respondents were collected (66 % female and 34% males), 18 years old and older (divided into three age groups). Participants were classified into two groups for each of the three subordinate scales of the NARS. Low negative attitude towards robots with a score below the mean of 3.4 or high negative attitude towards robots for above mean scores [73]. Table 7 summarizes the participants' information.

Table 7: Summary of participants' information

| | | | |
|---|---|---|---|
| Total | | N=110 | 100% |
| Gender | Male | N=37 | 33.6% |
| | Female | N=73 | 66.4% |

| | | | | | |
|---|---|---|---|---|---|
| Age | 18-35 | | | N=36 | 32.7% |
| | 36-55 | | | N=38 | 34.6% |
| | 56 + | | | N=36 | 32.7% |
| TAP | Very low | | | N=0 | 0% |
| | Low | | | N=19 | 17.3% |
| | Moderate | | | N=44 | 40% |
| | High | | | N=36 | 32.7% |
| | Very high | | | N=11 | 10% |
| NARS | Negative attitude toward situations of interaction with robots | | Low | N=75 | 68.2% |
| | | | High | N=35 | 31.8% |
| | Negative attitude toward the social influence of robots | | Low | N=38 | 34.6% |
| | | | High | N=72 | 65.4% |
| | Negative attitude toward emotions in interaction with robots | | Low | N=46 | 61.8% |
| | | | High | N=64 | 58.2% |

### 4.4.4 Results

*Descriptive and Chi-Square analysis.* Most of the participants preferred a V-shape structure or rectangle over all three others. Although a chi-square test of independence ($X^2$(n)=XX, P=XX) showed that there was no significant association between gender and the selection of body structure, results showed that male participants tend to prefer V-shape over all other forms (46%), while within the female participants the most frequently selected structure was the rectangle (29%) followed by the hourglass structure and V-shape (22%, each). Most participants (59%) chose to use a rounded outline in their designs, a chi-square test of independence was performed to examine the relationship between gender and outline preferences. The relation between these variables was significant, $X^2$ (1, $N$ = 110) = 4.4447, $p$ = .035. Male participants were more likely to choose a rounded outline than female participants (males=73%, females=52%); a chi-square test of independence showed no significant association, $X^2$ (2, $N$ = 110) = 1.47, $p$ = .48, between Age and outline design preferences.

Gender was found to have no effect on color scheme selection, $X^2$ (2, $N$ = 110) = 0.252, $p$ = .88 and GUI selection, $X^2$ (1, $N$ = 110) = 0.052, $p$ = .82. ; the most frequently selected color scheme was the White and Blue combination (48%). Most of the participants preferred to use the icons screen (64%) rather than the face display regardless of their gender or age group. NARS score significantly affected participants' preferences regarding the robot's GUI, $X^2$ (1, $N$ = 110) =98.337, $p < .00001$; all participants with a low negative attitude toward the social influence of robots (S2) preferred the face display (100%), and most of the participants with a high S2 preferred the icons display (97%). TAP was found to have no effect on the participants' preferences.

**Fitting a Hierarchical Cluster Analysis**

Summarizing the descriptive results can bring us to the conclusion that an optimal design would be a rounded, white-and-blue V-shape robot with an icons screen display, but this may be oversimplistic and nonrepresentative of the respondents' preferences. We turned to further data-driven analysis, utilizing a hierarchical cluster analysis to identify alternative clusters and detect how the robotic qualities group. Since all of our variables are nominal (categorical) and the common methods for unsupervised learning (cluster analysis) are suitable for continuous variables, choosing a method suitable for nominal variables is necessary. We used the nomclust function from the nomclust package for R. This function runs hierarchical cluster analysis (HCA) with objects characterized by nominal (categorical) variables. In our analysis, we used the combination of the LIN index to measure similarity along with the mean linkage method as recommended in Sulc and Rezankova, 2019 [74]. Since both evaluation criteria, PSFM and PSFE, suggested working with 3 clusters (as noted in the Dendrogram, see appendix A), we continued with 3 clusters for the analysis.

Attempts to characterize the three clusters with the help of the four characteristics: Body Structure, Outline shape, Color Scheme, and GUI had brought us to the following clusters: Cluster (1) is a white rounded A shape with a face display, cluster (2) is a white and blue rounded V shape with icons display, and cluster (3) is a white and blue, chamfered rectangle with a face display. The mosaic plot (each rectangle height represents the percentage within a column (cluster)) in Figure 13 illustrates the characterization of the three clusters.

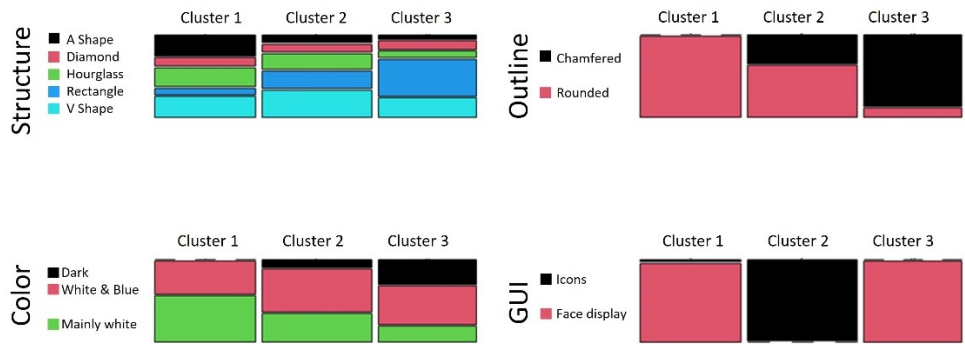

Figure 4: Characterization of the three clusters by VQs of the SAR and its GUI.

Descriptive statistics for examining the relationships between demographic variables of age and gender, and cluster affiliation, show that gender has no relation. All three clusters contain members of the three age groups, but cluster (2) has more participants at the ages of 36-55 and fewer participants at the ages of 56 and above compared to the two other clusters. The mosaic plot in figure 14 illustrates the association between clusters' affiliation and the demographic variables.

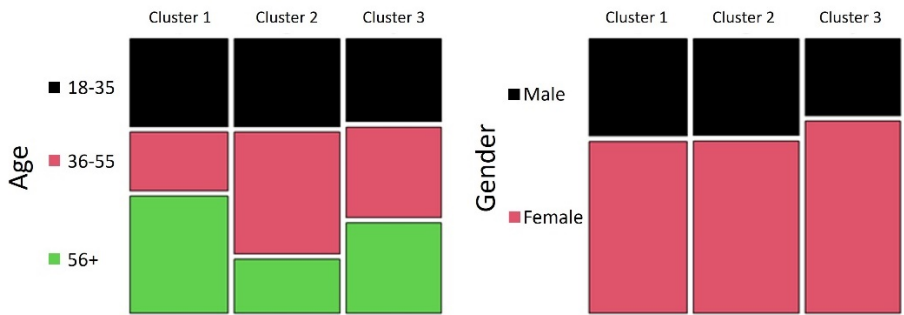

Figure 5: The association between clusters' affiliation and demographic variables

Descriptive statistics for examining the relationships between personality variables, S1: "Negative Attitude toward Situations of Interaction with Robots," S2: "Negative Attitude toward Social Influence of Robots," and S3: "Negative Attitude toward Emotions in Interaction with Robots." and cluster affiliation revealed that (S1) and (S2) have no relationship to the cluster affiliation. (S3) affects cluster affiliation; clusters 1 and 3 are composed of mainly participants with high levels of negative attitude toward emotions in interaction with robots. Cluster 2 is composed of high and low levels with a slight tendency to a low level. The mosaic plot in Figure 15 illustrates the association between clusters' affiliation and personality variables.

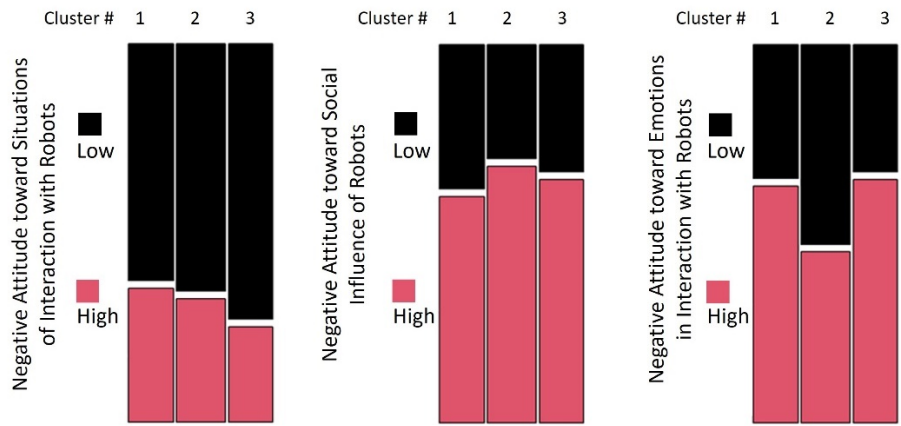

Figure 6: Examination of the association between clusters' affiliation and personality variables

### 4.4.5 *Summary*

We have found differences in participants' preferences regarding the design of SARs for home use. Some were related to gender, and some to participants' negative attitudes toward the social influence of robots. Participants assembled their robotic model by choosing its body structure, outline, and color scheme (out of 30 possible options); they also selected their preferred GUI. Exploring the chosen design outcomes revealed that 24 out of the 30 possible combinations (80%) were selected by at least two respondents. Two varieties share the first place as the most popular robot design with 11 selections (10% of the respondents). The two differ in all three VQs; the first is built of a chamfered-edged rectangle using a white and blue color scheme, and the second is a mainly white, rounded V-shape. Although it may seem that these results show no tendency toward a specific VQ, it was found that most participants choose VQs that are perceived as friendlier over those that are perceived as more elegant or innovative. This finding can be insightful for designers on what to incorporate in their future design work. The results demonstrate the importance of involving the users in the design process, as noted by the HCA.

## 5 DISCUSSION

Most research in the field of SARs design focuses either on evaluating existing off-the-shelf SARs [3,34,75-77] or on using participatory design processes to develop a new solution [78,79]. Among the studies that were using purposely designed SARs, the majority focused on anthropomorphism [9]. Very few studies dealt with isolated visual features using designated SARs [24].

To begin our research, we first conducted market research to explore and analyze SARs' VQs; we then deconstructed the design models into small building boxes for our studies. Our aim was to evaluate the effect of isolated VQs on user perception and explore users' preferences. We selected three basic VQs: body structure, outline, and color scheme in the context of personal assistant robots in the home domain. These three VQs were selected as they are related to users' first impressions of the SAR. Based on the literature review, we hypothesized that these VQs would significantly affect users' perception, as found in other design fields. We assessed users' design preferences and perceptions of these three VQs utilizing our own building boxes and the ability to design the robots' features from scratch. Based on two studies with a total of 270 participants (study 1, N=160, study 2, N=110), we demonstrate how VQs such as structure, outline, and color significantly affect users' perception of the SAR's characteristics and support hypotheses [H1a], [H1b], and [H1c]. These findings correlate with previous studies assessing the effect of structure, outline, and colors in different products [37,41].

In addition, we found that preference differences relate to users' gender and NARS score; these findings support hypothesis [H2]. Hypothesis [H1d] assumed that participants would choose VQs that share similar characteristics, and Hypothesis [H3] assumed that positive perceptions would correlate with design preferences. However, the results did not support these two assumptions. We found that out of 30 possible combinations for the SAR's design (excluding the GUI), 24 models (80%) were selected at least twice (no combination was selected only once). The two most popular varieties received 11 selections each (10% of the respondents). This implies there is no consensus among users regarding the appropriate appearance for SARs in the domestic environment, which may be caused by a lack of familiarity and experience with SARs.

We added the option of selecting the GUI in order to give the users an option to make the design more human-like if they wished to do so since anthropomorphism level is one of the most frequent categorizations of robots and was found to affect users' behaviors and attitudes in various contexts [80]. Our literature and market survey revealed that manufacturers often use the screen as the robot's head to increase its anthropomorphism [55]. Yet, on the other hand, users tend to prefer less anthropomorphized robots in different contexts [32-36]. Hence, we found it interesting to explore users' preferences when designing a new robot. The results were consistent with previous studies; most participants preferred to use the icons screen (64% n=70) rather than the face display. The participants' preferences regarding the robot's GUI were mostly affected by their NARS score; all participants with a low negative attitude toward the social influence of robots (S2) preferred the face display, and most of the participants with a high S2 preferred the icons display. These insights may suggest that using a screen that can perform as a face in some cases (but not always) is important to gain acceptance and reduce anxiety in some populations.

Analyzing the results of Study II (see Figure 13) brought us to the definition of three distinct clusters; each is suitable for a particular user group. We then cross-checked the results of the two studies to find the perceived characteristics for each cluster. Table 8 summarizes our three clusters of robots and their potential users. Note that each of the three clusters has at least one character perceived as friendly in the first study, implying humans seek this specific character when designing SARs for the home environment. This correlates with previous studies that investigated SARs' personalities and users' preferences; in a systematic review by Broadbent et al. [81], it was found that participants seek more sociable robots. The perception of SAR's sociability may be caused by its behavior [82], its appearance, or its voice [83].

Table 8: Three clusters of robots and their potential users

|  | Cluster 1 | Cluster 2 | Cluster 3 |
|---|---|---|---|
| **Body Shape** | A-shape | V-Shape | Rectangle |
| **Outline** | Rounded | Rounded | Chamfered |
| **Color Scheme** | Mainly white | White and blue | White and blue |
| **GUI** | Icons | Face | Icons |

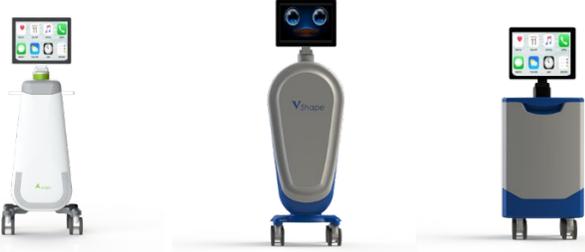

| | | | |
|---|---|---|---|
| **Users' perception** | Friendly | Threatening | Old fashioned |
| **(According to study 1)** | Innovative | Massive | Friendly |
|  | Elegant | Innovative |  |
|  | Medical | Friendly |  |
| **Gender** | Female/ Male | Female/ Male | Female |
| **Age** | 56 and above | 36-55 | 18 and above |
| **S1** | Low | Low | Low |
| **S2** | High | High | High |
| **S3** | High | Low | High |

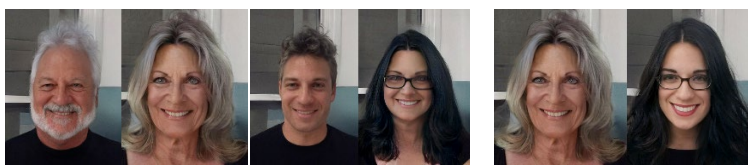

\* All images are a synthetic representation of the same features in various ages/genders. We used our first author's facial image to generate all images.

## 6   CONCLUSION, LIMITATIONS, AND FUTURE WORK

This paper presents an evaluation of the effect of three basic VQs: body structure, outline, and color on users' perception of the SAR's characteristics. Based on SARs market research and three preliminary studies, we created our own self-designed 30 SARs models using CAD software. These designs share most of their visual features (the same height, screen dimension and orientation, and wheels dimensions and location) and are divided by their body structure, outline, and color scheme. This allowed us to evaluate the effect of each isolated VQ using online questionnaires.

This paper offers empirical findings on how isolated VQs impact people's perception of a robot's characteristics: friendly, childish, innovative, threatening, old-fashioned, massive, elegant, medical, and gender. The results imply that to achieve user acceptance, designers must select VQs appropriate to the robot's role and its desired characteristics; moreover, users' preferences may vary by different personal factors. Hence, allowing the user to further apply minor design adjustments may increase their satisfaction. These conclusions, of course, can be relevant to more types of robots and products than SARs, particularly the ones we built here, but we find them extremely important, especially in the field of SAR design, as SARs aim to create social interactions.

We plan to continue deconstructing SARs' VQs to expand our knowledge and contribution in order to set up design guidelines to assist designers in the design process of a new SAR. In addition, the investigated VQs should be further evaluated using different scales and figures of SARs, as these may affect the results.

A key limitation of our study is the fact that participants watched the images of the different SARs online. We believe that the effect of the robot appearance should also be tested with real interaction rather than online questionnaires; therefore, our following studies will evaluate the effect of SARs' VQs on user behaviors involving live interaction and a real task. The nature of the study may also be the cause for the users' preferences diversity; although we defined the context of a personal assistant robot for home use, without a real interaction, participants may have interpreted this description differently according to their needs, desires, and presumptions. Another limitation may be due to the language of the study; studies were conducted in Israel among Hebrew-speaking participants and via snowball recruitment. To make sure that the perception describing words' translation to English has the same meaning, we followed a two-way translation procedure. However, chances exist that our results are sensitive to language or cultural differences.

To align these findings with our proposed relationships models [14], our subsequent studies will focus on evaluating user expectations in each context of use. What would be the relevant and desired characteristics of different SARs (by domains, e.g., healthcare, companion, educational, etc.)? These studies will explore different stakeholders' perceptions as well as the effect of culture. These findings are expected to support the design process of new SARs depending on their context of use, their intended role, and their users. And form design guidelines for future SARs

## 7 ACKNOWLEDGMENTS

This research was supported by the Ministry of Science Fund, grant agreement number: 81059, and by the ABC Foundation of Ben-Gurion University of the Negev through the Helmsley Charitable Trust, the Agricultural, Biological and Cognitive Robotics Initiative, and the George Shrut Chair in human performance management.

## REFERENCES

[1] Feil-Seifer, D., & Mataric, M. J. (2005, June). Defining socially assistive robotics. In *9th International Conference on Rehabilitation Robotics, 2005. ICORR 2005.* (pp. 465-468). IEEE.

[2] Matarić, M. J., & Scassellati, B. (2016). Socially assistive robotics. *Springer handbook of robotics*, 1973-1994.

[3] Lazar, A., Thompson, H. J., Piper, A. M., & Demiris, G. (2016, June). Rethinking the design of robotic pets for older adults. In Proceedings of the 2016 ACM Conference on Designing Interactive Systems (pp. 1034-1046).

[4] IFR (2020). Executive Summary World Robotics 2020 Service Robots. Int. Fed. Robot. Available at: https://ifr.org/img/worldrobotics/Executive_Summary_WR_2020_Service_Robots.pdf [Accessed May 5, 2021].

[5] Tavakoli, M., Carriere, J., & Torabi, A. (2020). Robotics, smart wearable technologies, and autonomous intelligent systems for healthcare during the COVID-19 pandemic: An analysis of the state of the art and future vision. Advanced Intelligent Systems, 2000071.

[6] Honig, S. S., Oron-Gilad, T., Zaichyk, H., Sarne-Fleischmann, V., Olatunji, S., & Edan, Y. (2018). Toward socially aware person-following robots. IEEE Transactions on Cognitive and Developmental Systems, 10(4), 936-954.

[7] Schulz, T. W., Herstad, J., & Tørresen, J. (2018). Moving with Style: Classifying Human and Robot Movement at Home. In *International Conferences on Advances in Computer-Human Interactions ACHI* (pp. 188-193). International Academy, Research and Industry Association (IARIA).

[8] Beer, J. M., Prakash, A., Mitzner, T. L., & Rogers, W. A. (2011). Understanding robot acceptance. Georgia Institute of Technology.

[9] Goetz, J., Kiesler, S., & Powers, A. (2003, November). Matching robot appearance and behavior to tasks to improve human-robot cooperation. In The 12th IEEE International Workshop on Robot and Human Interactive Communication, 2003. Proceedings. ROMAN 2003. (pp. 55-60). Ieee.

[10] Crilly, N., Moultrie, J., & Clarkson, P. J. (2004). Seeing things: consumer response to the visual domain in product design. Design studies, 25(6), 547-577.

[11] Kristoffersson, A., Severinson Eklundh, K. & Loutfi, A. Measuring the Quality of Interaction in Mobile Robotic Telepresence: A Pilot's Perspective. Int J of Soc Robotics 5, 89–101 (2013).

[12] Hoffman, G., & Ju, W. (2014). Designing robots with movement in mind. Journal of Human-Robot Interaction, 3(1), 91-122.

[13] Chatterjee, S., Parmet, Y., & Oron-Gilad, T. (2021). Body Language for Personal Robot Arm Assistant. International Journal of Social Robotics, 1-23.

[14] Liberman-Pincu, E., Van Grondelle, E. D., & Oron-Gilad, T. (2021, March). Designing Robots with Relationships in Mind: Suggesting Two Models of Human-socially Assistive Robot (SAR) Relationship. In Companion of the 2021 ACM/IEEE International Conference on Human-Robot Interaction (pp. 555-558).

[15] Onnasch, L., & Roesler, E. (2020). A Taxonomy to Structure and Analyze Human–Robot Interaction. International Journal of Social Robotics, 1-17.

[16] Broadbent, E., Kumar, V., Li, X., Sollers 3rd, J., Stafford, R. Q., MacDonald, B. A., & Wegner, D. M. (2013). Robots with display screens: a robot with a more human-like face display is perceived to have more mind and a better personality. PloS one, 8(8), e72589.

[17] Kalegina, A., Schroeder, G., Allchin, A., Berlin, K., & Cakmak, M. (2018, February). Characterizing the design space of rendered robot faces. In Proceedings of the 2018 ACM/IEEE International Conference on Human-Robot Interaction (pp. 96-104).

[18] Złotowski, J., Proudfoot, D., Yogeeswaran, K., & Bartneck, C. (2015). Anthropomorphism: opportunities and challenges in human–robot interaction. International journal of social robotics, 7(3), 347-360.

[19] Tung, F. W. (2011, July). Influence of gender and age on the attitudes of children towards humanoid robots. In International Conference on Human-Computer Interaction (pp. 637-646). Springer, Berlin, Heidelberg.

[20] Dautenhahn, K., Nehaniv, C. L., Walters, M. L., Robins, B., Kose-Bagci, H., Assif, N., & Blow, M. (2009). KASPAR–a minimally expressive humanoid robot for human–robot interaction research. Applied Bionics and Biomechanics, 6(3, 4), 369-397.

[21] Bartneck, C., & Forlizzi, J. (2004, September). A design-centred framework for social human-robot interaction. In RO-MAN 2004. 13th IEEE international workshop on robot and human interactive communication (IEEE Catalog No. 04TH8759) (pp. 591-594). IEEE.

[22] Blaurock, M., Čaić, M., Okan, M., & Henkel, A. P. A transdisciplinary review and framework of consumer interactions with embodied social robots: Design, delegate, deploy. International Journal of Consumer Studies.

[23] DiSalvo, C. F., Gemperle, F., Forlizzi, J., & Kiesler, S. (2002, June). All robots are not created equal: the design and perception of humanoid robot heads. In *Proceedings of the 4th conference on Designing interactive systems: processes, practices, methods, and techniques* (pp. 321-326).

[24] Björklund, L. (2018). Knock on Wood: Does Material Choice Change the Social Perception of Robots?.

[25] Tuch, A. N., Presslaber, E. E., StöCklin, M., Opwis, K., & Bargas-Avila, J. A. (2012). The role of visual complexity and prototypicality regarding

first impression of websites: Working towards understanding aesthetic judgments. International journal of human-computer studies, 70(11), 794-811.

[26] Conway, C.M., Pelet, J.E., Papadopoulou, P. and Limayem, M., 2010. Coloring in the Lines: Using Color to Change the Perception of Quality in E-Commerce sites. In ICIS (p. 224).

[27] Auster, C. J., & Mansbach, C. S. (2012). The gender marketing of toys: An analysis of color and type of toy on the Disney store website. Sex Roles, 67(7-8), 375-388.

[28] Benedek, J., & Miner, T (2002). Measuring Desirability: New methods for evaluating desirability in a usability lab setting.

[29] Lohse, M., Hegel, F., Swadzba, A., Rohlfing, K., Wachsmuth, S., & Wrede, B. (2007, February). What can I do for you? Appearance and application of robots. In Proceedings of AISB (Vol. 7, pp. 121-126).

[30] Li, D., Rau, P. L., & Li, Y. (2010). A cross-cultural study: Effect of robot appearance and task. International Journal of Social Robotics, 2(2), 175-186.

[31] Trovato, G., Cuellar, F., & Nishimura, M. (2016, November). Introducing 'theomorphic robots'. In 2016 IEEE-RAS 16th International Conference on Humanoid Robots (Humanoids) (pp. 1245-1250). IEEE.

[32] Piçarra, N., Giger, J. C., Pochwatko, G., & Możaryn, J. (2016). Designing social robots for interaction at work: Socio-cognitive factors underlying intention to work with social robots. Journal of Automation Mobile Robotics and Intelligent Systems, 10.

[33] Forlizzi, J., DiSalvo, C., & Gemperle, F. (2004). Assistive robotics and an ecology of elders living independently in their homes. Human–Computer Interaction, 19(1-2), 25-59.

[34] Wu, Y.H., Fassert, C. and Rigaud, A.S., 2012. Designing robots for the elderly: appearance issue and beyond. Archives of gerontology and geriatrics, 54(1), pp.121-126.

[35] Dario, P., Guglielmelli, E., Laschi, C., & Teti, G. (1999). MOVAID: a personal robot in everyday life of disabled and elderly people. Technology and Disability, 10(2), 77-93.

[36] Arras, K. O., & Cerqui, D. (2005). Do we want to share our lives and bodies with robots? A 2000 people survey: a 2000-people survey. Technical report, 605.

[37] Demirbilek, O., & Sener, B. (2003). Product design, semantics and emotional response. Ergonomics, 46(13-14), 1346-1360.

[38] Mugge, R., Govers, P. C., & Schoormans, J. P. (2009). The development and testing of a product personality scale. Design Studies, 30(3), 287-302.)

[39] Khalaj, J., & Pedgley, O. (2019). A semantic discontinuity detection (SDD) method for comparing designers' product expressions with users' product impressions. Design Studies, 62, 36-67.,

[40] Janlert, L. E., & Stolterman, E. (1997). The character of things. Design Studies, 18(3), 297-314.

[41] Mata, Marta Perez, Saeema Ahmed-Kristensen, Per Brunn Brockhoff, and Hideyoshi Yanagisawa. "Investigating the influence of product perception and geometric features." Research in Engineering Design 28, no. 3 (2017): 357-379.

[42] Otterbacher, Jahna, and Michael Talias. "S/he's too Warm/Agentic! The Influence of Gender on Uncanny Reactions to Robots." In 2017 12th ACM/IEEE International Conference on Human-Robot Interaction (HRI, pp. 214-223. IEEE, 2017.

[43] Bernotat, J., Eyssel, F., & Sachse, J. (2017, November). Shape it–the influence of robot body shape on gender perception in robots. In International Conference on Social Robotics (pp. 75-84). Springer, Cham.

[44] Tama, D., & Öndoğan, Z. (2014). Fitting evaluation of pattern making systems according to female body shapes. Fibres & Textiles in Eastern Europe.

[45] Stroessner, S. J., Benitez, J., Perez, M. A., Wyman, A. B., Carpinella, C. M., & Johnson, K. L. (2020). What's in a shape? Evidence of gender category associations with basic forms. Journal of Experimental Social Psychology, 87, 103915.

[46] Fogelström, E. (2013). Investigation of shapes and colours as elements of character design. Uppsala Universitet

[47] Larson, C. L., Aronoff, J., & Steuer, E. L. (2012). Simple geometric shapes are implicitly associated with affective value. Motivation and Emotion, 36(3), 404-413.

[48] Toet, A., & Tak, S. (2013). Look out, there is a triangle behind you! The effect of primitive geometric shapes on perceived facial dominance. i-Perception, 4(1), 53-56.

[49] Salgado-Montejo, Alejandro, Tapia Leon, Isabell, Elliot, Andrew J, Salgado, Carlos José, & Spence, Charles. (2015). Smiles over Frowns: When Curved Lines Influence Product Preference. Psychology & Marketing, 32(7), 771–781. https://doi.org/10.1002/mar.20817

[50] Bertamini, M., Palumbo, L., Gheorghes, T. N., & Galatsidas, M. (2016). Do observers like curvature or do they dislike angularity?. British Journal of Psychology, 107(1), 154-178.

[51] Bar, M., & Neta, M. (2006). Humans prefer curved visual objects. Psychological science, 17(8), 645-648.

[52] Seckler, M., Opwis, K., & Tuch, A. N. (2015). Linking objective design factors with subjective aesthetics: An experimental study on how structure and color of websites affect the facets of users' visual aesthetic perception. Computers in Human Behavior, 49, 375-389.

[53] Taft, C. (1997). Color meaning and context: Comparisons of semantic ratings of colors on samples and objects. Color Research & Application: Endorsed by Inter-Society Color Council, The Colour Group (Great Britain), Canadian Society for Color, Color Science Association of Japan, Dutch Society for the Study of Color, The Swedish Colour Centre Foundation, Colour Society of Australia, Centre Français de la Couleur, 22(1), 40-50.

[54] Taylor, C., Clifford, A., & Franklin, A. (2013). Color preferences are not universal. Journal of Experimental Psychology: General, 142(4), 1015.

[55] E. Kalegina, A., Schroeder, G., Allchin, A., Berlin, K., & Cakmak, M. (2018, February). Characterizing the design space of rendered robot faces. In Proceedings of the 2018 ACM/IEEE International Conference on Human-Robot Interaction (pp. 96-104).

[56] Powers, A., & Kiesler, S. (2006, March). The advisor robot: tracing people's mental model from a robot's physical attributes. In Proceedings of the 1st ACM SIGCHI/SIGART conference on Human-robot interaction (pp. 218-225).

[57] Diana, C., & Thomaz, A. L. (2011). The shape of simon: creative design of a humanoid robot shell. In CHI'11 Extended Abstracts on Human Factors in Computing Systems (pp. 283-298).

[58] Ferstl, Y., Kokkinara, E., & McDonnell, R. (2016, July). Do I trust you, abstract creature? A study on personality perception of abstract virtual faces. In Proceedings of the ACM Symposium on Applied Perception (pp. 39-43).

[59] Powers, A., & Kiesler, S. (2006, March). The advisor robot: tracing people's mental model from a robot's physical attributes. In Proceedings of the 1st ACM SIGCHI/SIGART conference on Human-robot interaction (pp. 218-225).

[60] Diana, C., & Thomaz, A. L. (2011). The shape of simon: creative design of a humanoid robot shell. In CHI'11 Extended Abstracts on Human Factors in Computing Systems (pp. 283-298).

[61] Ferstl, Y., Kokkinara, E., & McDonnell, R. (2016, July). Do I trust you, abstract creature? A study on personality perception of abstract virtual faces. In Proceedings of the ACM Symposium on Applied Perception (pp. 39-43).

[62] Nomura, T., Suzuki, T., Kanda, T., & Kato, K. (2006). Measurement of negative attitudes toward robots. Interaction Studies, 7(3), 437-454.

[63] Ratchford, M., & Barnhart, M. (2012). Development and validation of the technology adoption propensity (TAP) index. Journal of Business Research, 65(8), 1209-1215.

[64] Bartneck, C., Kulić, D., Croft, E., & Zoghbi, S. (2009). Measurement instruments for the anthropomorphism, animacy, likeability, perceived intelligence, and perceived safety of robots. International journal of social robotics, 1(1), 71-81.

[65] Carpinella, C. M., Wyman, A. B., Perez, M. A., & Stroessner, S. J. (2017, March). The Robotic Social Attributes Scale (RoSAS) Development and Validation. In Proceedings of the 2017 ACM/IEEE International Conference on human-robot interaction (pp. 254-262).

[66] Liberman-Pincu, E., & Oron-Gilad, T. (2021, August). Impacting the perception of socially assistive robots-evaluating the effect of visual qualities among children. In *2021 30th IEEE International Conference on Robot & Human Interactive Communication (RO-MAN)* (pp. 612-618). IEEE.

[67] Liberman-Pincu, E. (2021, March). Audrey-Flower-like Social Assistive Robot: Taking Care of Older Adults in Times of Social Isolation during the COVID-19 Pandemic. In Companion of the 2021 ACM/IEEE International Conference on Human-Robot Interaction (pp. 613-614).

[68] Hsiao, K. A., & Chen, L. L. (2006). Fundamental dimensions of affective responses to product shapes. International Journal of Industrial Ergonomics, 36(6), 553-564.

[69] Liberman-Pincu, E., & Bitan, Y. (2021). FULE—Functionality, Usability, Look-and-Feel and Evaluation Novel User-Centered Product Design Methodology—Illustrated in the Case of an Autonomous Medical Device. Applied Sciences, 11(3), 985.

[70] Putnam, C., Kolko, B., & Wood, S. (2012, March). Communicating about users in ICTD: leveraging HCI personas. In Proceedings of the Fifth International Conference on Information and Communication Technologies and Development (pp. 338-349).

[71] Schrum, M. L., Johnson, M., Ghuy, M., & Gombolay, M. C. (2020, March). Four years in review: Statistical practices of Likert scales in human-robot interaction studies. In *Companion of the 2020 ACM/IEEE International Conference on Human-Robot Interaction* (pp. 43-52).

[72] Castro, J. W., Acuña, S. T., & Juristo, N. (2008, October). Integrating the personas technique into the requirements analysis activity. In 2008 Mexican International Conference on Computer Science (pp. 104-112). IEEE.

[73] Cramer, H., Kemper, N., Amin, A., Wielinga, B., & Evers, V. (2009). 'Give me a hug': the effects of touch and autonomy on people's responses to embodied social agents. *Computer Animation and Virtual Worlds*, *20*(2-3), 437-445.

[74] Šulc, Z., & Řezanková, H. (2019). Comparison of similarity measures for categorical data in hierarchical clustering. Journal of Classification, 36(1), 58-72.

[75] Pino, M., Boulay, M., Jouen, F., & Rigaud, A. S. (2015). "Are we ready for robots that care for us?" Attitudes and opinions of older adults toward socially assistive robots. Frontiers in aging neuroscience, 7, 141.

[76] von der Pütten, A., & Krämer, N. (2012, March). A survey on robot appearances. In 2012 7th ACM/IEEE International Conference on Human-Robot Interaction (HRI) (pp. 267-268). IEEE.

[77] Reeves, B., & Hancock, J. (2020). Social robots are like real people: First impressions, attributes, and stereotyping of social robots. Technology, Mind, and Behavior, 1(1).

[78] McGinn, C., Bourke, E., Murtagh, A., Donovan, C., Lynch, P., Cullinan, M. F., & Kelly, K. (2020). Meet Stevie: A Socially Assistive Robot Developed Through Application of a 'Design-Thinking'Approach. Journal of Intelligent & Robotic Systems, 98(1), 39-58.

[79] Hegel, F., Eyssel, F., & Wrede, B. (2010, September). The social robot 'flobi': Key concepts of industrial design. In 19th International Symposium in Robot and Human Interactive Communication (pp. 107-112). IEEE.

[80] Akdim, K., Belanche, D., & Flavián, M. (2021). Attitudes toward service robots: analyses of explicit and implicit attitudes based on anthropomorphism and construal level theory. *International Journal of Contemporary Hospitality Management*.

[81] Broadbent, E., Stafford, R., & MacDonald, B. (2009). Acceptance of healthcare robots for the older population: review and future directions. International journal of social robotics, 1(4), 319.

[82] Walters, M. L., Lohse, M., Hanheide, M., Wrede, B., Syrdal, D. S., Koay, K. L., ... & Severinson-Eklundh, K. (2011). Evaluating the robot personality and verbal behavior of domestic robots using video-based studies. Advanced Robotics, 25(18), 2233-2254.

[83] Powers, A., & Kiesler, S. (2006, March). The advisor robot: tracing people's mental model from a robot's physical attributes. In Proceedings of the 1st ACM SIGCHI/SIGART conference on Human-robot interaction (pp. 218-225).

**APPENDICES**

**Appendix A: Three clusters dendrogram of user preferences derived from the nomclust hierarchical cluster analysis.**

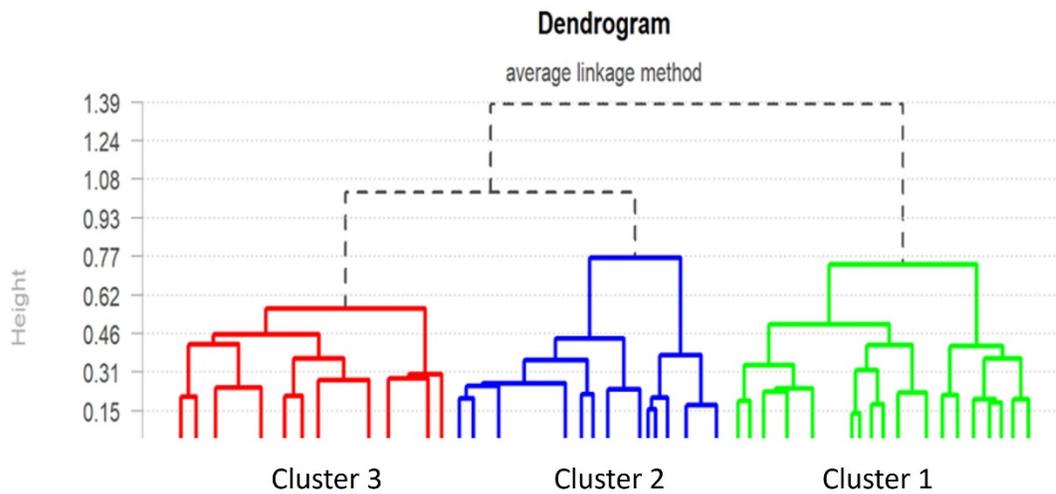